\documentclass[aps,prl,twocolumn,groupedaddress]{revtex4}
\usepackage{color}
\usepackage{ifthen}
\usepackage{amsmath,amssymb}
\usepackage{graphicx}
\usepackage{wrapfig}
\usepackage{subfig}
\usepackage{booktabs}
\usepackage{dcolumn}

\newcommand\be{\begin{equation}}
\newcommand\ee{\end{equation}}
\newcommand\ie{\emph{i.\,e.\ }}

\newcommand\Kt[1]{\tilde\kappa\ifthenelse{\equal{#1}{}}{}{({#1})}}

\newcommand\Zggt[2]{
  \mathcal{\tilde{Z}}
  \ifthenelse
  {\equal{#1}{}}
  {}
  {({#1},{#2})}
} 
\newcommand\Zgt[2]{{{\tilde{Z}}_{#1}\ifthenelse{\equal{#2}{}}{}{({#2})}}} 

\begin{document}


\title{McGenus: A Monte Carlo algorithm to predict RNA secondary structures with pseudoknots}


\author{Micha\"el Bon $^{1}$, Cristian Micheletti $^{2}$
and Henri Orland $^{1}$
}

\address{(1) Institut de Physique Th\'eorique, CEA Saclay, CNRS URA 2306, 91191 Gif-sur-Yvette, France\\
(2) SISSA, Scuola Internazionale Superiore di Studi Avanzati and CNR-IOM Democritos, Via Bonomea 265, I-34136 Trieste, Italy}


\begin{abstract}
{We present McGenus, an algorithm to predict RNA secondary structures
with pseudoknots. The method is based on a classification of RNA
structures according to their topological genus. McGenus can treat
sequences of up to 1000 bases and performs an advanced stochastic search
of their minimum free energy structure allowing for non trivial
pseudoknot topologies.  Specifically, McGenus employs a Monte Carlo algorithm with replica exchange for minimizing a general scoring function which includes
not only free energy contributions for pair stacking, loop penalties,
etc. but also a phenomenological penalty for the genus of the pairing graph.  The good performance of
the stochastic search strategy was successfully validated
against TT2NE which uses the same free energy parametrization and performs  exhaustive or partially
exhaustive structure search,  albeit for much shorter sequences (up to
200 bases).  Next, the method was applied to other RNA sets, including an extensive tmRNA database, yielding results that are competitive with existing algorithms. Finally, it is shown that McGenus highlights possible limitations in the free energy scoring function. The algorithm is available as a web-server at http://ipht.cea.fr/rna/mcgenus.php.\\
\phantom{xx}\\
DOI: 10.1093/nar/gks1204}
 \end{abstract}

\maketitle



\section{Introduction}
In the past twenty years, there has been a tremendous increase of
interest in RNA by the biological community. This biopolymer,
which was at first merely considered as a simple information
carrier, was gradually proven to be a major actor in the biology
of the cell \cite{book}.

Since the RNA functionality is mostly determined by its three-dimensional
conformation, the accurate prediction of RNA folding from
the nucleotide sequence is a central
issue \cite{Tinocco1991}.
It is strongly believed that the biological activity of RNA (be it enzymatic or regulatory), is implemented through the binding of 
some unpaired bases of the RNA with their ligand. It is thus crucial to have a precise and reliable map of all the pairings taking place in RNA and to correctly identify loops.
The complete list of all Watson-Crick and Wobble base pairs in RNA is called the {\em secondary structure} of RNA.

In this paper, we stick to the standard assumption that there is an effective free energy which
governs the formation of secondary structures, so that the optimal folding of an RNA sequence
is found as the minimum free energy structure (MFE for short). The
problem of finding the MFE structure given a certain sequence has been
conceptually solved provided the MFE is planar, \ie the MFE structure
contains no pair ($i$,$j$), ($k$,$l$) such that $i<k<j<l$ or $k<i<l<j$. In that
case, polynomial algorithms which can treat long RNAs assuming a
mostly linear free energy model have been proposed
\cite{nussinov,zuker0,mccaskill}. Otherwise, the MFE structure is said
to contain pseudoknots and finding it has been shown to be an
NP-complete problem with respect to the sequence length
\cite{lyngso2}.

In a previous paper \cite{TT2NE}, we proposed an algorithm, TT2NE, which consists in searching for the exact MFE structure for a certain form of the energy function, where pseudoknots are penalized according to a topological index, namely their genus.
TT2NE relies on the ``maximum weighted independent set" (WIS) formalism. In this approach, an RNA structure is viewed as a collection of stem-like structures (helices possibly comprising bulges of size 1 or internal loops of size $1 \times 1$), called ``helipoints" \cite{TT2NE}, defined in the next section. 
Given a certain sequence, the set of all possible
helipoints is enumerated and used to build a weighted graph. The graph vertices are the helipoints and their weight is given by -1 times the helipoint  free energy. Two vertices are linked by an arc if and only if the corresponding helipoints are not compatible in the same secondary structure. Incompatibilities arise, for example, when two helipoints share one or more bases as this could imply the formation of base triplets, which is forbidden.
Finding the MFE structure thus amounts to finding the maximum weighted independent set of the graph, \ie the set of pairwise compatible helipoints for which the overall free energy is minimum.

Both McGenus and TT2NE utilize the same energy function, defined in terms of helipoints and genus penalty as well as the same initial  graph.
The difference between the two lies in the search algorithm for the MFE. 
While in TT2NE the secondary structure is built by adding or removing helipoints in a deterministic order, in McGenus, they are added or removed one at a time according to a stochastic Monte Carlo Metropolis scheme. As in TT2NE, there is no restriction on the pseudoknot topology that McGenus can generate.  A server implementation of McGenus can be found at http://ipht.cea.fr/rna/mcgenus.php.

In the following and in the numerical implementation of McGenus, we will restrict ourselves to the energy function and genus penalty described in detail in \cite{TT2NE}. While in TT2NE, the energy form was dictated by the requirement to allow for a branch and bound procedure, here in McGenus we insist that there is no such restriction on the form of the  energy function. It can for instance include loop and pseudoknot entropies. Furthermore, the penalty for pseudoknots needs not be proportional to the genus as in TT2NE, but may depend also on the topology of each individual pseudoknot (see below). Therefore, by modifying the energy function, it is possible to improve on the results that we will present below.
As stated in the introduction, the initial graph is generated in the same way as in \cite{TT2NE}.

\section{Materials and methods}

In the present framework, the folded structure of a given RNA sequence
is given by the set of helipoints which minimizes the free energy. 
We recall that a helipoint is \emph{an ensemble} of helices (defined as a stack of base pairs possibly comprising bulges of size 1 or internal loops of size $1 \times 1$) that are demarcated by the same extremal (initial and terminal) base pairs. Given two extremal pairs $(i,j)$ and $(k,l)$, the set $\omega^{ij}_{kl}$ of all helices that end with these two pairs can be generated and their individual energies calculated according to a given energy model. The free energy $\Delta F^{ij}_{kl}$ of the helipoint is then computed as
\begin{eqnarray}
&\exp&{(-\beta \Delta F^{ij}_{kl})} = \sum_{\substack{h \in \omega^{ij}_{kl}}} \exp{(-\beta e(h))} \\  
&&\mbox{with} \ \beta = (k_B T)^{-1} \nonumber
\end{eqnarray}
where $e(h)$ is the free energy of formation of helix $h$.
In our implementation, to speed up the computation of this sum, helices of non-negative (\ie unfavorable) energies are neglected, since their Boltzmann weight would strongly suppress their contribution. Helipoints are stem-like structural building blocks which account for all possible internal pairing possibilities that occur between their extremal pairs.
We shall denote by $\{h_1, ... , h_N\}$ the set of all helipoints that can
possibly arise from the pairings of nucleotides in the given sequence
(their total number $N$, is clearly sequence dependent). 
We stress that the set of enumerated helipoints comprises all possible helipoints, and hence is not restricted to maximal ones.

Clearly, a given RNA structure $S$ is fully specified by a collection of compatible helipoints. It is therefore convenient to identify $S$ with a binary vector, $\vec{\sigma}^S$, of length $N$ and whose $i$-th component, $\sigma^S_i$ takes on the value 0 or 1 according to whether helipoint $h_i$ belongs to $S$. The free energy of $S$ can accordingly be written as:

\begin{equation}
F_S = \sum_{i=1}^{N} \sigma^S_i\, \Delta F(h_{i}) + \mu \, g (S) \ .
\label{eqn:free_en}
\end{equation}

The first term is the additive contribution of the free energy $\Delta F$ of individual helipoints, and is parametrized as in
\cite{TT2NE}. The second term weights the topological complexity of the
structure, measured by its genus $g$ \cite{orland,vernizzi}. Unlike the first term which is local, 
the genus, which is a
non-negative integer, depends globally on all the helipoints. The parameter $\mu \ge 0$ is used to penalize
structures with excessively large values of the genus, in agreement
with the phenomenological observation that the genus of most
naturally-occurring RNA structures of size up to 600 bases, is smaller than 4. Based on
previous studies \cite{TT2NE},  the default value of the genus penalty $\mu$ is
set equal to 1.5 kcal/mol.

It is implicitly assumed that the free energy of incompatible sets of
helipoints is infinite.

\subsection{Advanced Monte Carlo search of MFE structures}

The minimization of the free energy (\ref{eqn:free_en}) is
carried out by a Monte Carlo (MC) exploration of structure space, 
that is over the set of
possible $\vec{\sigma}$ vectors.
Starting from a structure $S$ where only one helipoint is present, at each
Monte Carlo step, one of the helipoints $h_i$ is added ($\sigma_i=0 \to \sigma_i=1$) or removed 
($\sigma_i=1 \to \sigma_i=0$). The helipoint to be modified is picked with a
biased probability favoring the addition (resp. removal) of helipoints with low (resp. high)
free energy $e$. The biasing is inspired by the heat-bath MC
algorithm. Specifically, the {\em a priori} probability to pick helipoint $h_i$ to be changed in structure $S$
is given by:
\begin{equation}
w^S_i = {\sigma_i^S + (1 - \sigma^S_i) \, e^{-\beta \Delta F(h_i)}  \over \sum^\prime_{j=1..N} \,  \sigma_j^S + (1 - \sigma^S_j) \, e^{-\beta \Delta F(h_j)}} 
\label{eqn:forward}
\end{equation}
where the prime superscript indicates that helipoints incompatible with $S$ are not considered. Changing the state of $h_i$ defines a trial structure, $S^\prime$, which is accepted with probability
\begin{equation}
\min\left[1,{ w^{S^\prime}_i \over w^S_i} e^{-\beta (F_{S^\prime} - F_S)}\right]\ .
\end{equation}
\noindent The above acceptance criterion is a generalization of the standard Metropolis rule  and ensures that, in the long run, the generated structures are sampled with probability given by the canonical weight $\exp[-\beta F_S]$.

The stochastic generation of structures is carried
out within a Monte Carlo algorithm with replica exchange where several simulations
are run in parallel at different inverse temperatures $\beta$. The values of $\beta$ are chosen so as to cover a range  
of thermal energies ${1 / \beta}$, going from about one tenth of the smallest helipoint energy up to 
the largest helipoint energy. At regular time
intervals, swaps are proposed between structures at neighboring
temperatures and are accepted with the generalized Metropolis
criterion described in ref.~\cite{Orlandini:1998:IMA}. 
The Markov replicas at the lowest temperature progressively populate
structures of low free-energy, and a record is kept of the lowest
energy structures which are finally provided as output.

Finally, we point out that the Monte Carlo optimization can be performed not only within the whole space of secondary structures (unconstrained search) but is straightforwardly restricted to topologically-constrained subspaces. In particular, by introducing {\em{ad hoc}} ``infinite" energy penalties in eq.~\ref{eqn:free_en}, the search can be restricted to structures whose genus, topology or extent of pseudoknots satisfy some preassigned constraints.
The web-server interface allows the user to set such thresholds, e.g. to account for knowledge based constraints.

\subsection{Generalized Topological Penalties}
\begin{figure}[h]
\begin{center}
\includegraphics[scale=0.2]{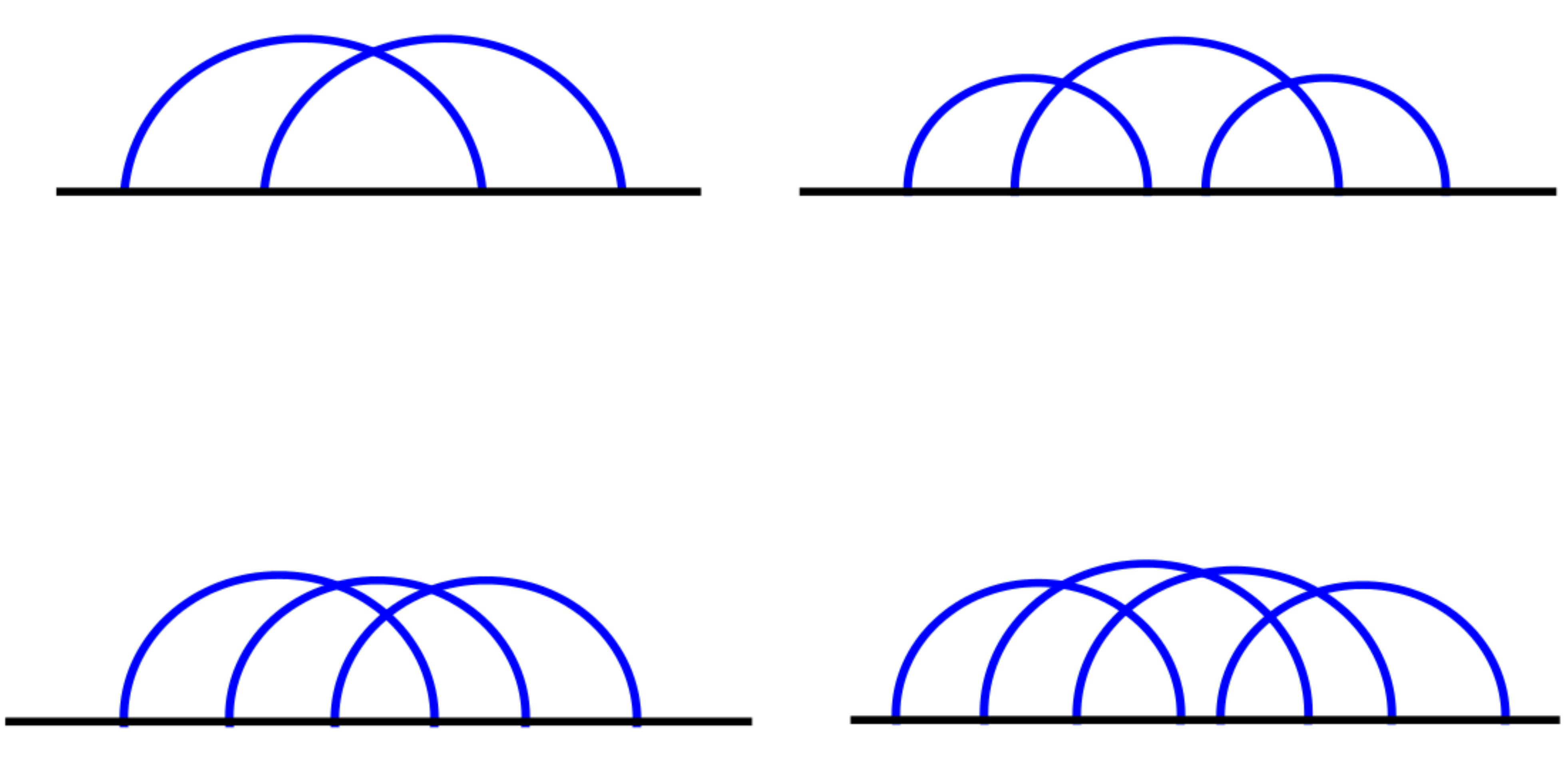}
\end{center}
\caption{\emph{The only four primitive pseudoknots of genus 1 \cite{bon}. } }
\label{fig1}
\end{figure}

As we have previously reported \cite{bon,RMT}, any RNA complex pseudoknot
structure may be built from of a set of building blocks, called
primitive pseudoknots.  A pseudoknot is termed primitive if it is
{\em (i)} irreducible, {\em{i.e.}} its standard diagrammatic representation cannot be disconnected by cutting one
backbone line and {\em (ii)} contains no nested pseudoknot, that is it
cannot be disconnected by cutting two backbone lines, see
Fig.~\ref{fig1}. An arbitrary pseudoknotted structure can be decomposed in a collection of primitive pseudoknots and its total genus is the sum of the genii of its primitive constituents \cite{bon}.

Therefore, it makes sense to assign different penalties to pseudoknots having same
{genus} but with different primitive components.  For example,
all tmRNAs have total genus 3 or 4 and contain no primitive
pseudoknots of {genus} larger than 1. In the present implementation, we
propose only two options: i) we forbid primitive pseudoknots of {genus}
larger than 1 (by assigning them an infinite penalty) but the overall
structure can have any total genus or ii) we assign a global penalty
proportional to the total genus and do not take into account the
decomposition of the structure into primitive blocks.

\section{RESULTS AND DISCUSSION}

We have carried out an extensive comparison of McGenus predictions against those of other methods. For this purpose we used hundreds of RNA sequences from various sets, including: the dataset previously used for TT2NE  \cite{TT2NE}, an extensive set of tmRNAs \cite{tmrna} and the more limited set of pseudoknotted RNA molecules for which the structural data is available in the protein databank (PDB).
Over such diverse datasets, the predictive performance is aptly conveyed by the {\em sensitivity} of the method, that is the fraction of
pairs in the reference (native) structure that are correctly predicted by the method. Depending on the context we shall also report on the positive predicted value (PPV). The PPV corresponds to the fraction of predicted pairs that are found in the native structure, and hence measures the incidence of false positives in the predicted contacts. We have considered this measure for the PDB set, but not for the tmRNA set whose entries, often corresponding to putative native structures derived from homology, are known to potentially lack several native contacts, as in the paradigmatic case of Aste.yell.\_TRW-322098\_1-426 \cite{tmrna}. A visual representation of this structure can be found in the RNA STRAND database \cite{rnastrand} under the reference TMR\_00037.

From an overall point of view, the tests are aimed at elucidating two issues that are central to any MFE-based method. The first issue, regards the algorithmic effectiveness of the energy minimization, while the second regards the viability of the energy parametrization within the considered space of secondary structures.
 The former is most clearly ascertained by comparing algorithms employing the same energy parametrization. This step is crucial for the second aspect too. In fact, the appropriateness or the limitations of a given energy parametrization and/or of the considered secondary structure space, can be exposed in a non-ambiguous way only if the minimization algorithm is well-performing.

Following the above-mentioned logical order, we started by comparing the predictions of McGenus against TT2NE on a database of 47 short sequences ($<$ 209 bases) used in \cite{TT2NE}. Because McGenus and TT2NE rely on the same 
energy parametrization\cite{mathews1999}, the comparison provides a stringent test of the effectiveness of the energy-minimization procedure. In fact, we recall that TT2NE is based on an exhaustive, or nearly exhaustive search in sequence space.
Despite the stochastic, non-exhaustive and much faster McGenus searches, its performance turned out to be optimal. Over the full data set, McGenus returned exactly the same MFE structures as TT2NE, as well as all the suboptimal structures.

To extend the assessment of McGenus minimization performance for longer chains, that cannot be addressed by TT2NE, we considered UNAFold \cite{zuker0}, a MFE-based algorithm restricted to secondary structures without pseudoknots. We used a customized version of UNAFold which employs the same energy parametrization as McGenus. However, it cannot yet be compared to McGenus since it outputs secondary structures in terms of base pairs rather than helipoints. To circumvent this difficulty, we generated all the lowest lying secondary structures (within 1kCal/mol from the lowest energy structure) using the algorithm presented in ref.~\cite{wuchty}. To match the description of the structure in terms of helipoints, we made clusters of secondary structures sharing the same extremities of their helical fragments. We then resummed them (in terms of their Boltzmann weights) and as a result the energy discrepancy between the two approaches is negligible. In the sequel, we will refer to this process as cUNAFold.

The comparison was carried out over the complete set of 590 sequences of genus 3, 4 or 5 from the tmRNA database \cite{tmrna} with lengths in the 200-500 range. 
To assess the efficiency of the minimization algorithm of McGenus, we ran it over our sample of 590 sequences, with the constraint $g_{max}=0$ and compared it with the output of cUNAFold. The average MFE from McGenus with $g_{max}=0$  is -105.1 kCal/mol while that of cUNAFold is -106.7 kCal/mol. Interestingly enough, out of the 590 sequences, 191 sequences are predicted to have identical secondary structures by both algorithms. This comparison shows the good efficiency of McGenus minimization algorithm.

In the non-zero genus case, for each of the 590 sequences, McGenus returned structures with lower free energy than cUNAFold. 
On the average, the free energy of the McGenus predicted structures was -125 kcal/mol.

These two tests prove the effectiveness of the energy-minimization scheme adopted by McGenus and we accordingly turned our attention to the overall predictive performance of the method (sensitivity).
For this purpose we used again the 590 sequences of genus 3, 4 or 5 from the tmRNA database \cite{tmrna} and compared McGenus predictions against {McQfold \cite{metzler}, HotKnots \cite{ren}, ProbKnot \cite{prob} and UNAFold \cite{UNAFold} on this set. We did not compare McGenus against PKnots \cite{eddy} and gfold \cite{reidys}, as the original articles claim that they cannot handle sequences longer than 200 bases.} We recall that UNAFold predictions are restricted to secondary structures free of pseudoknots, while ProbKnot and McQfold can output any topology of pseudoknot. The genus of each of McGenus prediction was enforced not to exceed the genus of the native structures of the dataset. As discussed in \cite{TT2NE}, the setting of the corresponding parameter $g_{max}$ can be decided by the user. In this report, for each test sequence, we chose to set $g_{max}$ to the appropriate, native, value to illustrate the performance of McGenus when it is driven in the appropriate secondary structure search space.

The total number of base pairs to be predicted in the set is 56740. {The UNAFold,  McQfold, ProbKnot, HotKnots and McGenus arithmetic averages of the sensitivity over all sequences are respectively 37\%, 42\%, 43\%, 39\% and 43\%, with a respective standard deviation of 14\%, 15\%, 14\%, 14\% and 16\%. A closer look at the secondary structures output by ProbKnot and HotKnots showed that none of them contained any pseudoknot}. Therefore the performance of McGenus is not inferior to that of the few methods that can handle sequences of comparable length. {Even without resorting to advanced comparative tests  \cite{Mat1,Con1}, the consistent sensitivity of these 5 algorithms allows
to conclude that their performance is very similar.}

The fact that the average sensitivity of the five methods is below 50\% poses the question of whether it can be improved by tweaking the energy parameters or by suitably further constraining the space of secondary structures over which the minimization is performed.
We focus on the latter aspect as the first has been already discussed in \cite{TT2NE}. The space of secondary structures considered by prediction schemes based on abstract, graph-theoretical representations, include structures that are unphysical, {\em{i.e.}} that cannot be realized in a three-dimensional space because of chain connectivity constraints.

The impact of this major difficulty can be lessened by excluding from further considerations those structures that present physically-unviable or atypical levels of entanglement. To illustrate this point, we note that, in the mentioned dataset of 590 molecules, only H-pseudoknots which span less than 70 bases are present. By enforcing such knowledge-based constraint on the search space, the sensitivity of McGenus is boosted from 43\% to 53\% with a standard deviation of 18\%. To assess the statistical significance of this improvement, we performed the Welch t-test. We find a t-value of $t=10$, which with a total of 1168 degrees of freedom implies a $p-$value smaller than $10^{-7}$, i.e. the improvement is definitely significant.

Introducing the constraint in structure space clearly results in higher energies for the predicted structures. In fact the average free energy was -125 kcal/mol without the constraint while it is -114 kcal/mol with the restriction of the pseudoknot length.
Notwithstanding the reduction of the search space due to the pseudoknot-length constraint, the structures returned by McGenus have an energy that is significantly lower than the reference, (putative) native structures, which is about  -~73kcal/mol.
The free energy difference appears too large to be accounted for by the neglected contribution of loop entropy, missing chain-connectivity constraints or imperfect parametrization of the potentials, which are well established. A more plausible source of discrepancy could the missing contacts in the homology-derived native structure of the tmRNA database.

To check this last point, we have studied the unconstrained version McGenus on a set of 4 sequences from the protein databank (PDB) { with $g_{max}$ being fixed to the native genus}.
{Their PDB ids are: 
1Y0Q (length=229, g=1), 3EOH (length=412, g=1),  2A64 (length=417, g=1) and 2H0W (length=151, g=2). The structures of these entries are unambiguously known from X-ray scattering data and contain very few long and non-hybridized RNA sequences ({\em{i.e.}} not bound to proteins, DNA or other molecules). Accordingly, the McGenus performance on this set was higher than for the tmRNA set. The sensitivity for 1Y0Q, 3EOH,  2A64 and 2H0W was equal to  87\%, 39\%, 50\% and 72\%, respectively while the PPV was equal to 90\%, 38\%, 35\% and 84\%, respectively. Again, the structures predicted by McGenus have a lower free energy than the native ones. This indicates that, besides accounting for topological effects, further improvements of secondary structure predictions would probably require a better parametrization of the free energy. The generality and flexibility of the McGenus search algorithm ought to allow for incorporating any such modifications in a transparent way.}

{Finally let us discuss the choice of a maximum genus. Ideally, one should perform the computation with a completely unconstrained genus. However, there are two difficulties to this approach.} {First, since steric constraints are only limitedly accounted for by available pseudoknot prediction algorithms (including McGenus), the predicted structures can  be sterically impossible and hence associated to an excessively high genus. Secondly, the computational time required to explore the unrestricted genus space could be impractical. To overcome these difficulties and restrict the search space one can profitably introduce knowledge-based constraints. In particular, the statistical PDB analysis of ref.~\cite{bon} provides a quantitative indication for the dependence of the genus on the length of naturally-occurring RNA sequences. The data can be clearly used to provide a phenomenological upper bound to $g_{max}$.}
{Alternatively, a user could explore a few different increasing values of $g_{max}$  and carry out a supervised evaluation of the results by taking into account (i) the phenomenological constraints and (ii) the possibility that structures with excessively large genus value are returned because of the imperfect treatment of steric constraints.}

{To illustrate this last point, we ran McGenus on a set of 792 5S rRNA sequences of length around 150, with no pseudoknot. We set $g_{max}=3$ which according to the study of ref.~\cite{bon} {(see Fig.~10 therein)} is very large. The number of sequences predicted with genus 0 (i.e. without pseudoknots) is 258, with genus 1 is 500, with genus 2 is 34 and with genus 3 is 0.} {Consistently with the remarks made in the context of H-pseudoknots, the results indicate that performance of pseudoknot prediction algorithms could certainly benefit by improving the current handling of chain connectivity and excluded volume constraints.}

\subsection{CPU time}

The CPU time required by McGenus to fold an RNA sequence depends on the total number of Monte Carlo steps.
For a tm-RNA of length 400, the typical number of helipoints is 3500. For each sequence, we use 10 replicas, and overall 3000 $\times$ number of helipoints steps to achieve these results. The result is typically returned in 15 minutes on a parallel quadcore computer { (Intel Xeon CPU @2.66GHz). The current implementation of McGenus on the server is not parallelized.}

\section{Conclusion}
In this article, we presented McGenus, an efficient algorithm for RNA pseudoknot prediction, which proves
that classifying pseudoknots according to their genus is a relevant
and successful concept. {We showed that on a set of RNA structures from the tm-RNA database \cite{tmrna}, McGenus allows treatment of sequences
of sizes up to 1000 with a typical CPU time of 15 minutes for a 500 long sequence on a quadcore CPU}, with a performance {that is comparable or better than the few methods that  can treat sequences with comparable length.}

In order to further improve the performance of McGenus, we see 3 main directions:
I) improvement on the computing techniques, in particular on the parallelization of the algorithm.
II) improvement of the functional form and parametrization of the energy model (likely to have an impact also on the parametrization of pseudoknot-free methods such as UNAFold). 
III) inclusion of steric constraints.
 
\section{Acknowledgements}
We acknowledge financial support from the Italian Ministry of research, grant FIRB - Futuro in Ricerca N. RBFR102PY5. The authors wish to thank A. Capdepon for setting up the McGenus server at http://ipht.cea.fr/rna/mcgenus.php. 

\bibliographystyle{unsrt}

\end{document}